\documentclass[12pt]{article}
\pdfoutput=1

\usepackage{amsmath}
\usepackage{amssymb}
\usepackage{epsfig}
\usepackage{epstopdf}
\usepackage{latexsym}
\usepackage{color}
\numberwithin{equation}{section}
\usepackage[
      colorlinks=false,
      linkcolor=darkblue,  
      urlcolor=blue,    
      filecolor=blue,     
      citecolor=red,
linktocpage=true,
      pdfstartview=FitV,
      bookmarksopen=true    
      ]{hyperref}

\begin{document}

\begin{titlepage}

\centerline
\centerline
\centerline
\bigskip
\bigskip
\centerline{\Huge \rm A gravity interpretation for the complex}
\bigskip
\centerline{\Huge \rm saddles of the ABJM and spindle indices}
\bigskip
\bigskip
\bigskip
\bigskip
\bigskip
\bigskip
\bigskip
\centerline{\rm Minwoo Suh}
\bigskip
\centerline{\it School of General Education, Kumoh National Institute of Technology,}
\centerline{\it Gumi, 39177, Korea}
\bigskip
\centerline{\tt minwoosuh1@gmail.com} 
\bigskip
\bigskip
\bigskip
\bigskip
\bigskip
\bigskip
\bigskip

\begin{abstract}
\noindent The superconformal index of the ABJM theory diverges exponentially by the contribution of the terms from the shifts in complex chemical potentials. In order to resolve the puzzle, we study the stability condition of probe M5-brane in Kerr-Newman black hole. The condition is used to examine the stability of the solutions with the shifts in chemical potentials. We also obtain the stability conditions for the spindle and the $AdS_7$ black holes.
\end{abstract}

\vskip 5cm

\flushleft {December, 2026}

\end{titlepage}

\tableofcontents

\section{Introduction and summary}

In the last decade, via the AdS/CFT correspondence, \cite{Maldacena:1997re}, microstates of supersymmetric $AdS$ black holes have been successfully counted by the partition functions of dual field theories. In particular, for magnetically charged static black holes, the Bekenstein-Hawking entropy was reproduced by the topologically twisted index of dual field theories, \cite{Benini:2015eyy, Benini:2015noa}. Furthermore, by introducing complex chemical potentials and taking sufficient limits, $e.g.$, the Cardy-like limit or large $N$ limit of the superconformal index, \cite{Romelsberger:2005eg, Kinney:2005ej}, the entropy of electrically charged rotating black holes was accounted, \cite{Cabo-Bizet:2018ehj, Choi:2018hmj, Benini:2018ywd}. More precisely, the superconformal index is a grand-canonical partition function that counts the 1/16-BPS states in the theory, and its Legendre transform with respect to reduced chemical potentials agrees with the black hole entropy, \cite{Hosseini:2017mds, Hosseini:2018dob, Choi:2018fdc}.

For electrically charged rotating black holes in $AdS_5$, however, the superconformal index of $\mathcal{N}=4$ super Yang-Mills theory (SYM) appears to allow shifts of chemical potentials, and the contribution of the shifted terms diverges exponentially, \cite{Aharony:2021zkr}. See also \cite{Mamroud:2022msu, Aharony:2024ntg}. The gravitational on-shell action, which gives the minus of the Bekenstein-Hawking entropy, also has the same problem. However, when considering a particular D3-brane wrapping an $S^3$ inside the $S^5$ and an $S^1$ inside the $AdS_5$, the solutions corresponding to the shifts in the on-shell action turned out to make positive contributions to the brane action without bound, $i.e.$, the configuration is unstable. Thus, one has to exclude the solutions that correspond to shifts in the on-shell action. In the field theory, this also excludes the shifts that result in the superconformal index diverging. \footnote{See \cite{BenettiGenolini:2026hmz} for recent derivation of the brane action through equivariant localization. See also \cite{Gautason:2025per}.}

The identical puzzle was raised for the superconformal index of ABJM theory in \cite{BenettiGenolini:2023rkq}. See $e.g.$, \cite{Choi:2019zpz, Bobev:2019zmz, Nian:2019pxj, GonzalezLezcano:2022hcf} and \cite{Bobev:2022wem, Bobev:2024mqw, Bobev:2025ltz} for the previous studies on the ABJM index. The ABJM index diverges by the contribution of the terms from the shifts in complex chemical potentials. In this paper, in order to resolve this puzzle, we study probe M5-brane wrapping an $S^5$ inside the $S^7$ and an $S^1$ inside the $AdS_4$ in Kerr-Newman black holes. For non-diverging on-shell action of the probe M5-brane, we find
\begin{equation}
\text{Im}\left(S_\text{M5}\right)\,>\,0\,.
\end{equation}
We show that the minimal solutions without the shifts in chemical potentials are always stable under this condition. For the multi-charged solutions with the shifted chemical potentials, we find an analogous condition to the $\mathcal{N}=4$ SYM index which appeared to match the $\{1,N,r\}$ solution to the Bethe Ansatz equations, \cite{Aharony:2021zkr}. It is an open question to find the corresponding solution from the ABJM index which explains the condition.

Furthermore, we calculate the probe M5-brane action for accelerating and rotating black holes, \cite{Ferrero:2020twa, Cassani:2021dwa}, which have a horizon geometry of weighted projective space, $\mathbb{WCP}^1_{[n_-,n_+]}$, known as a spindle, $\Sigma$. For the non-accelerating case, the accelerating black hole reduces to the Kerr-Newman black hole. We also study the probe M2-brane action in rotating black holes in $AdS_7$. Detailed stability analysis of the probe brane on-shell actions for the spindle and the $AdS_7$ black holes are open problems.

There are several interesting avenues to explore. First, the contribution of the probe D3-brane action was identified to the non-perturbative correction to the $\mathcal{N}=4$ SYM index, \cite{Aharony:2021zkr}. Analogously, the contribution of the probe M5-brane would reproduce the non-perturbative correction to the ABJM index, \cite{BenettiGenolini:2023rkq}, and to the spindle index, \cite{Inglese:2023wky, Inglese:2023tyc, Colombo:2024mts}. Second, it would be interesting to study probe branes in the black hole background as a defect, as it was studied in the $AdS_5\times{S}^5$ case, \cite{Chen:2023lzq, Cabo-Bizet:2023ejm}. Third, it seems important to understand the relation of the probe brane stability condition and the Kontsevich-Segal-Witten (KSW) criterion for the allowability of complex metrics in the context of the gravitational path integral, \cite{Kontsevich:2021dmb, Witten:2021nzp}. Recently, the KSW criterion was applied to electrically charged rotating black holes in $AdS$, \cite{BenettiGenolini:2025jwe, BenettiGenolini:2026raa, Krishna:2026rma}. Fourth, we considered rotating $AdS_4$ black holes with two out of four independent electric charges. Although it would be most interesting to study the black hole solutions with four independent charges, only the BPS solutions of such kind are known, \cite{Hristov:2019mqp}. Fifth, studying probe brane action of the $AdS_6$ rotating black holes, \cite{Chow:2008ip}, for the stability analysis is an open problem.

In section \ref{sec2}, we review Kerr-Newman black holes in $AdS_4$, their BPS limit and thermodynamics, following \cite{BenettiGenolini:2023rkq}. We construct the probe M5-brane and calculate the on-shell action. We study the stability condition for the probe M5-brane. In section \ref{sec3}, we obtain the probe M5-brane action for accelerating and rotating black holes. In appendix \ref{appA}, we obtain the probe M5-brane action for $AdS_4$ black holes with a rotation and two electric charges. In appendix \ref{appB}, we obtain the probe M2-brane action for rotating black holes in $AdS_7$. In appendix \ref{appC}, for the non-accelerating case, we show that the probe M5-brane action for accelerating black holes reduces to the one for Kerr-Newman black holes.

\section{Rotating black holes} \label{sec2}

This section is a review of the Kerr-Newman black holes in four-dimensional minimal gauged supergravity. In particular, following the approach of \cite{Cabo-Bizet:2018ehj, Cassani:2019mms}, we consider their BPS limit and thermodynamics. Via the AdS/CFT correspondence, \cite{Maldacena:1997re}, the on-shell action reproduces the partition function of the dual ABJM theory. We closely follow the analysis presented in \cite{BenettiGenolini:2023rkq}. 

\subsection{Black hole solutions}

The action of four-dimensional minimal gauged supergravity, \cite{Freedman:1976aw}, is
\begin{equation} \label{minact}
S\,=\,\frac{1}{16\pi{G}_4}\int_{Y_4}\left(\mathcal{R}+6-\mathcal{F}^2\right)\text{vol}_\mathcal{G}\,,
\end{equation}
where $\mathcal{R}$ is the Ricci scalar, $\mathcal{F}=d\mathcal{A}$ with the $U(1)$ gauge field, $\mathcal{A}$, and the cosmological constant, $\Lambda=-3$. Electrically charged rotating black hole solution, \cite{Carter:1968ks}, is given by
\begin{align} \label{minsol}
ds^2\,&=\,-\frac{\Delta_r\Delta_\theta}{B\Xi^2}dt^2+\sin^2\theta{B}\left(d\phi+a\Delta_\theta\frac{\Delta_r-\left(1+r^2\right)\left(r^2+a^2\right)}{BW\Xi^2}dt\right)^2 \notag \\
&+W\left(\frac{dr^2}{\Delta_r}+\frac{d\theta^2}{\Delta_\theta}\right)\,, \notag \\
\mathcal{A}\,&=\,\frac{mr\sinh\delta}{W\Xi}\Big(\Delta_\theta\,dt-a\sin^2\theta\,d\phi\Big)+\gamma\,dt\,,
\end{align}
where $\gamma$ is a constant, $\theta\in[0,\pi],\phi\in[0,2\pi)$, and
\begin{align} \label{minfs}
\Delta_r\,=&\,\left(r^2+a^2\right)\left(1+r^2\right)-2mr\cosh\delta+m^2\sinh^2\delta\,, \notag \\
\Delta_\theta\,=&\,1-a^2\cos^2\theta, \qquad W\,=\,r^2+a^2\cos^2\theta\,, \qquad \Xi\,=\,1-a^2\,, \notag \\
B\,\equiv&\,\frac{\Delta_\theta\left(r^2+a^2\right)^2-a^2\sin^2\theta\Delta_r}{W\Xi^2}\,.
\end{align}
The solution is in a frame that is non-rotating at infinity, \cite{Caldarelli:1999xj, Gibbons:2004ai}.

The largest root of $\Delta_r$, $r=r_+$, is where the outer horizon is. At constant $t$ and $r$ outside the horizon, the geometry is topologically a sphere and closes off at $\theta=0,\pi$ with $\phi\sim\phi+2\pi$. We analytically continue to a Euclidean solution by a Wick rotation, $t=-it_E$. At the horizon, the geometry caps off smoothly, only with the identification,
\begin{equation}
(t_\text{E},\phi)\,\sim\,(t_\text{E},\phi+2\pi)\,\sim\,(t_\text{E}+\beta,\phi-i\Omega\beta)\,,
\end{equation}
where the inverse temperature and angular velocity at the horizon are, respectively,
\begin{equation}
\beta\,=\,4\pi\frac{a^2+r_+^2}{\Delta'_r(r_+)}\,, \qquad \Omega\,=\,a\frac{1+r_+^2}{a^2+r_+^2}\,.
\end{equation}
There are two obvious commuting Killing vectors, $\partial_t$, $\partial_\phi$, and the surface, $\{r=r_+\}$, is a Killing horizon of
\begin{equation}
V\,=\,\frac{\partial}{\partial{t}}+\Omega\frac{\partial}{\partial\phi}\,.
\end{equation}
The electrostatic potential is
\begin{align}
\Phi_e\,&:=\,\imath_V\mathcal{A}|_{r=r_+}-\imath_V\mathcal{A}|_{r\rightarrow\infty} \notag \\
&=m\sinh\delta\frac{r_+}{a^2+r_+^2}\,,
\end{align}
where we fixed the gauge, $\gamma=-\Phi_e$. The Bekenstein-Hawking entropy at the horizon is
\begin{equation}
S\,=\,\frac{\pi}{G_4}\frac{r_+^2+a^2}{1-a^2}\,.
\end{equation}

We introduce a coordinate change, \cite{Hawking:1998kw},
\begin{equation}
\frac{\cos\Theta}{z}\,=\,r\cos\theta\,, \qquad z^{-2}\,=\,\frac{r^2\Delta_\theta+a^2\sin^2\theta}{\Xi}\,,
\end{equation}
and we find that the solution is an asymptotically anti-de Sitter space at $z\rightarrow0$,
\begin{align}
ds^2\,&\sim\,\frac{dz^2}{z^2}+\frac{1}{z^2}\left(-dt^2+\left(d\Theta^2+\sin^2\Theta\,d\phi^2\right)\right)\,, \notag \\
\mathcal{A}\,&\sim\,-\Phi_edt\,.
\end{align}
With $\widehat{\phi}=\phi+i\Omega{t}_\text{E}$, the boundary metric is 
\begin{equation}
ds_\text{bdry}^2\,=\,dt_\text{E}^2+\Big(d\theta^2+\sin^2\theta(d\widehat{\phi}-i\Omega{d}t_E)^2\Big)\,,
\end{equation}
and the identification is given by
\begin{equation}
(t_\text{E},\widehat{\phi})\,\sim\,(t_\text{E},\widehat{\phi}+2\pi)\,\sim\,(t_\text{E}+\beta,\widehat{\phi})\,.
\end{equation}

In the Euclidean signature, the on-shell action was computed by holographic renormalization, \cite{BenettiGenolini:2023rkq},
\begin{equation}
I\,=\,\frac{\beta}{2G_4\Xi}\left[m\cosh\delta-r_+\left(r_+^2+a^2\right)-\frac{r_+m^2\sinh^2\delta}{r_+^2+a^2}\right]\,.
\end{equation}
It satisfies the quantum statistical relation,
\begin{equation}
I\,=\,-S+\beta\left(E-\Omega\,J-\Phi_eQ_e\right)\,,
\end{equation}
where $E$, $J$, $Q_e$ are the energy, angular momentum, and electric charge of the black hole, respectively.

\subsection{Supersymmetry}

The supersymmetry condition for the black hole solutions, \eqref{minsol}, is, \cite{Kostelecky:1995ei, Caldarelli:1998hg},
\begin{equation} \label{susycmin}
E\,=\,J+Q_e\, \qquad \Leftrightarrow \qquad a\,=\,\coth\delta-1\,.
\end{equation}
With the supersymmetry condition, \eqref{susycmin}, $\Delta_r$ in \eqref{minfs} reduces to a sum of squares,
\begin{equation}
\Delta_r|_\text{SUSY}\,=\,\left(r^2-\coth\delta+1\right)^2+\coth^2\delta\left(r-m\frac{\sinh^2\delta}{\cosh\delta}\right)^2\,.
\end{equation}
Thus, if all parameters are real, we find
\begin{equation}
r_*^2\,=\,\coth\delta-1\,, \qquad m^2\,=\,\frac{\cosh^2\delta}{e^\delta\sinh^5\delta}\,,
\end{equation}
which fix the horizon radius, $r_*$, and $m$ by $\delta$.  We also find $\Delta'_r|_\text{SUSY}(r_*)=0$, and supersymmetry and regularity of the Lorentzian metric imply extremality.

For the Euclidean solutions, the metric is generically complex, \cite{BenettiGenolini:2023rkq}, and we have
\begin{equation}
m\,=\,\frac{1}{\sinh^2\delta}[r_+\cosh\delta\mp{i}\left(\sinh\delta\left(1+r_+^2\right)-\cosh\delta\right)]\,.
\end{equation}

For the supersymmetric solutions, the chemical potentials are
\begin{align}
\beta\,=&\,\mp2\pi{i}\frac{r_+^2+r_*^4}{\left(r_+^2-r_*^2\right)\left(1\mp2ir_++r_*^2\right)}\,, \notag \\
\Omega\,=&\,r_*^2\frac{1+r_+^2}{r_+^2+r_*^4}\,, \notag \\
\Phi_e\,=&\,r_+\frac{r_+r_*^2+r_+\mp{i}\left(r_+-r_*\right)\left(r_++r_*\right)}{r_+^2+r_*^4}\,,
\end{align}
and they satisfy
\begin{equation} \label{mconstr}
\beta\left(1-2\Phi_e+\Omega\right)\,=\,\pm2\pi{i}\,.
\end{equation}
The different signs refer to the two branches of solutions. We refer to the upper and lower signs as the first and second branches, respectively.

We obtain the BPS locus when extremality is imposed in addition to supersymmetry and is denoted by a subscript, $*$, as in \cite{Cassani:2019mms}. The extremal values of the chemical potentials are
\begin{equation}
\Omega_*\,=\,1\,, \qquad \Phi_{e*}\,=\,1\,,
\end{equation}
and the constraint in \eqref{mconstr} becomes meaningless. Thus, we define the ``reduced chemical potentials" for the supersymmetric solutions,
\begin{equation}
\tau_g\,\equiv\,\beta\frac{\Omega-\Omega_*}{2\pi{i}}\,, \qquad \varphi_g\,\equiv\,\beta\frac{\Phi_e-\Phi_{e*}}{2\pi{i}}\,,
\end{equation}
and we find
\begin{equation} \label{mtph}
\tau_g\,=\,\pm\frac{1-a}{1+a\mp2ir_+}\,,\qquad \varphi_g\,=\,\mp\frac{a\mp{i}r_+}{1+a\mp2ir_+}\,.
\end{equation}
They satisfy
\begin{equation} \label{tprel}
\tau_g-2\varphi_g\,=\,\pm1\,.
\end{equation}
The on-shell action is given by \footnote{This is distinct from the on-shell action for the $AdS_5$ case, where it is independent of the branches. See (4.29) of \cite{Aharony:2021zkr}.}
\begin{equation} \label{oasusy}
I|_\text{SUSY}\,=\,\mp\frac{\pi}{G_4}\frac{\varphi_g^2}{\tau_g}\,.
\end{equation}

\subsection{Comparison to field theory results}

The AdS/CFT dictionary of the four-dimensional Newton constant, $G_4$, and the rank of the gauge group of the dual ABJM theory, $N$, \cite{Aharony:2008ug}, is 
\begin{equation}
\frac{1}{G_4}\,=\,\frac{2\sqrt{2}}{3}N^\frac{3}{2}\,.
\end{equation}
The on-shell action can be expressed in terms of the field theory parameter, $\tau$,
\begin{equation}
I|_\text{SUSY}(\tau_g=\tau)\,=\,\mp\frac{\pi}{3\sqrt{2}}N^\frac{3}{2}\frac{\left(\tau\mp1\right)^2}{\tau}\,.
\end{equation}
This precisely matches the unrefined index of ABJM theory, \cite{BenettiGenolini:2023rkq}.

For a supersymmetric solution on a branch, we consider the shift of reduced chemical potentials, \cite{BenettiGenolini:2023rkq},
\begin{equation} \label{shifts}
\tau'_g\,=\,\tau_g+2n_\Omega\,, \qquad \varphi'_{g1}\,=\,\varphi_{g1}+2n_{e,1}\,, \qquad \varphi'_{g2}\,=\,\varphi_{g2}+2n_\Omega-2n_{e,1}\,,
\end{equation}
which leaves the solution on the same branch. The on-shell action for the shifted solution is
\begin{equation}
I|_\text{SUSY}\,=\,\mp\frac{2\sqrt{2}\pi}{3}N^\frac{3}{2}\frac{(\varphi_{g1}+2n_{e,1})(\varphi_{g2}+2n_\Omega-2n_{e,1})}{\tau_g+2n_\Omega}\,.
\end{equation}
With $n_\Omega=0$, the on-shell action diverges in the shift term,
\begin{equation}
\text{Re}(I|_\text{SUSY})\,=\,\pm{n}_{e,1}^2\frac{8\sqrt{2}\pi}{3}N^\frac{3}{2}\text{Re}\frac{1}{\tau_g}+\mathcal{O}(n_{e,1})\,.
\end{equation}
Thus, the contribution of $e^{-I|_\text{SUSY}}$ of the first/second branch would diverge for Re$(\tau_g)$ negative/positive, respectively. We will resolve this puzzle in the next section.

\subsection{Uplift to 11 dimensions}

We uplift the black hole solution of minimal gauged supergravity, \eqref{minsol}, to a solution of eleven-dimensional supergravity. We employ the uplift formula for $U(1)^4$-gauged $\mathcal{N}=2$ supergravity, \cite{Cvetic:1999xp}, by taking all four $U(1)$ gauge fields to be equal.

The eleven-dimensional metric is given by
\begin{equation} \label{meleven}
ds_{11}^2\,=\,ds_4^2+\frac{1}{g^2}\sum_{a=1}^4\left(d\mu_a^2+\mu_a^2\left(d\phi_a+g\mathcal{A}\right)^2\right)\,,
\end{equation}
where $ds_4^2$ is the four-dimensional metric and $\mu_1^2+\mu_2^2+\mu_3^2+\mu_4^2=1$.

The four-form flux is given by
\begin{equation}
F_{(4)}\,=\,2g\epsilon_{(4)}-\frac{1}{2g^2}\sum_{a=1}^4d\left(\mu_a^2\right)\wedge{d}\phi_a\wedge*_4\mathcal{F}\,,
\end{equation}
where $\epsilon_{(4)}$ is the volume form of the four-dimensional metric, $ds_4^2$, and $*_4$ is the Hodge dual with respect to that metric. From now on, we set $g=1/2$ to match the black hole solutions.

We calculate $F_{(7)}=*F_{(4)}$ and the six-form potential of $F_{(7)}=dC_{(6)}$ can be chosen to be
\begin{equation} \label{msixf}
C_{(6)}\,=\,8d\left(\mu_2^2\right)\wedge{d}\left(\mu_3^2\right)\wedge{d}\phi_2\wedge{d}\phi_3\wedge{d}\phi_4\wedge\mathcal{A}+\text{other\,\,terms}\,,
\end{equation}
where we only presented the term relevant to the discussion in the next section.

\subsection{The brane action}

Now we consider the M5-brane action at $r=r_+$, $\theta=\frac{\pi}{2}$, $\mu_1=0$, along the $\phi$ direction. The action of a M5-brane is given by, $e.g.$, \cite{Beccaria:2023cuo},
\begin{equation} \label{mdbi}
S_\text{M5}\,=\,-\frac{1}{\left(2\pi\right)^5\ell_p^6}\int\left(d^6x\sqrt{-\text{det}\left(g_\text{M5}\right)}\mp{P}[C_{(6)}]\right)\,,
\end{equation}
where $g_\text{M5}$ is the induced metric on the M5-brane worldvolume, $P[C_{(6)}]$ is the pull-back of the six-form potential to the worldvolume, and the $\pm$ sign refers to a brane/anti-brane. Via the AdS/CFT correspondence, the Planck length, $\ell_p$, is related to the rank of the dual ABJM theory, $N$, by, e.g., \cite{Choi:2018fdc},
\begin{equation}
\frac{R_{S^7}^6}{\left(2\pi\right)^5\ell_p^6}\,=\,\frac{N}{\pi^3}\,,
\end{equation}
with $R_{S^7}=2R_{AdS_4}$ and $R_{AdS_4}=1$. For the second term in the brane action, \eqref{mdbi}, only the term of the six-form potential presented in \eqref{msixf} contributes. It is a product of the integral of the integral of $8\mathcal{A}$ on the $S^1$ in $AdS_4$ and the integral of $d\left(\mu_2^2\right)\wedge{d}\left(\mu_3^2\right)\wedge{d}\phi_2\wedge{d}\phi_3\wedge{d}\phi_4$ on the $S^5$ in $S^7$. The first integral is
\begin{equation}
8\int_{S^1}\mathcal{A}\,=\,-\frac{8ma\sinh\delta}{r_+\left(1-a^2\right)}\int_{S^1}d\phi\,=\,-\frac{16\pi{m}a\sinh\delta}{r_+\left(1-a^2\right)}\,,
\end{equation}
and the second integral is $4\text{Vol}\left(S^5\right)=4\pi^3$.

Although the metric induced on the M5-brane from \eqref{meleven} is not diagonal, the determinant is the product of the determinant of the $S^5$ metric and of $g_{\phi\phi}$ in \eqref{minsol}. The integral of the determinant of $S^5$ with radius $2$ is $2^5\text{Vol}\left(S^5\right)=2^5\pi^3$. The component, $g_{\phi\phi}$, on the M5-brane is
\begin{equation}
g_{\phi\phi}\,=\,\left[\frac{r_+^2+a^2}{r_+\left(1-a^2\right)}\right]^2\,.
\end{equation}
Choosing a sign for the square root, we obtain
\begin{equation} \label{metterm}
\int{d}^6x\sqrt{-\text{det}\left(g_\text{M5}\right)}\,=\,-2^6\pi^4i\frac{r_+^2+a^2}{r_+\left(1-a^2\right)}\,.
\end{equation}
For the choice of the sign, if $a^2<1$ and $r_+$ is large, the Lorentzian black hole is causally well-behaved and the contribution of \eqref{metterm} to the path-integral measure, $\text{exp}(iS_\text{M5})$, is bounded in absolute value, \cite{Aharony:2021zkr}. The M5-brane action is
\begin{equation} \label{mM5act}
S_\text{M5}\,=\,-\frac{N}{\pi^3}\left[-\pi^4i\frac{r_+^2+a^2}{r_+\left(1-a^2\right)}\pm\pi^4\frac{ma\sinh\delta}{r_+\left(1-a^2\right)}\right]\,,
\end{equation}
where the $\mp$ sign implies a brane/anti-brane.

Furthermore, after imposing the supersymmetry condition, \eqref{susycmin}, the brane action simplifies and can be expressed in terms of the reduced chemical potentials,
\begin{equation} \label{bactf}
S_\text{M5}\,=\,\pm\pi{N}\frac{a\mp{i}r_+}{a-1}\,=\,\pm\pi{N}\frac{\varphi_g}{\tau_g}\,,
\end{equation}
where we employed \eqref{mtph} in the last step.

In appendix \ref{appA}, also for the black holes with two different charges, we have calculated the M5-brane action. From the result, we assume the M5-brane action for general chemical potentials, $(\tau_g,\varphi_{g,a})$, $a=1,2,3,4$, to be
\begin{align} \label{fourchargecond}
S_\text{M5}^a\,=&\,\pi{N}\frac{\varphi_{g,a}}{\tau_g}\,, \qquad {(1^\text{st}\,\,\text{branch})}\,, \\
S_\text{M5}^a\,=&\,-\pi{N}\frac{\varphi_{g,a}}{\tau_g}\,, \qquad {(2^\text{nd}\,\,\text{branch})}\,.
\end{align}

The Euclidean M5-brane contributes to the partition function by $\text{exp}\left(-I|_\text{SUSY}+iS_\text{M5}\right)$. The $i$ factor appears since the branes do not extend in the time direction and, thus, they do not get a factor $i$ when being Wick rotated to Euclidean time.

If the contribution of the M5-branes increases the partition function, an arbitrary number of M5-branes will increase the contribution without a bound, and it is unstable. Thus, only the background that decreases the contribution should be considered, and we have
\begin{equation} \label{mincond}
\text{Im}\left(S_\text{M5}\right)\,>\,0\,.
\end{equation}

\subsection{Stable gravity solution and the index} \label{sgs}

We start by considering the minimal solutions without the shifts. For the brane action, \eqref{bactf}, with \eqref{tprel} and find
\begin{equation}
S_\text{M5}\,=\,\pm\pi{N}\frac{\varphi_g}{\tau_g}\,=\,\pm\frac{\pi}{2}N\left(1\mp\frac{1}{\tau_g}\right)\,.
\end{equation}
To have positive imaginary part of the brane action, \eqref{mincond}, we find
\begin{equation}
\text{Im}[S_\text{M5}]\,=\,-\frac{\pi}{2}N\text{Im}\left[\frac{1}{\tau_g}\right]\,=\,\frac{\pi}{2}N\frac{\text{Im}[\tau_g]}{|\tau_g|^2}\,>\,0\,,
\end{equation}
and it reduces to
\begin{equation}
\text{Im}[\tau_g]\,>\,0\,.
\end{equation}
From \eqref{mtph} we find
\begin{equation}
\text{Im}[\tau_g]\,=\,\frac{2(1-a)r_+}{|1+a\mp2ir_+|^2}\,>\,0\,,
\end{equation}
and it is always positive as we have
\begin{equation}
a\,<\,1\,, \qquad r_+\,>\,0\,.
\end{equation}
Hence, the M5-brane probe is always stable for the minimal solutions.


Now, from the brane action for the two-charged solutions, \eqref{fourchargecond}, we study the stability with the shifted chemical potentials. Consider the stability conditions arising from a single M5-brane at $\mu_2=0$ and from the union of M5-branes at $\mu_1=0$. Using $\tau_g-\varphi_{g,1}-\varphi_{g,2}=1$ on the first branch, \eqref{condmult}, the two conditions are
\begin{equation}
\text{Im}\left(\frac{\varphi_{g,1}}{\tau_g}\right)>0\,, \qquad \text{Im}\left(\frac{-1-\varphi_{g,1}}{\tau_g}\right)>0\,,
\end{equation}
and hence
\begin{equation} \label{multicond}
\text{Im}\left(-\frac{1}{\tau_g}\right)>\text{Im}\left(\frac{\varphi_{g,1}}{\tau_g}\right)>0\,.
\end{equation}
As the gravitational chemical potentials are shifts of the ones of the index by even integers, $\varphi_{g,1}=\varphi_1+2n_{e,1}$ and $\tau_g=\tau+2n_\Omega$, they share the same imaginary part, and a shift of $\varphi_{g,1}$ by $2$ moves $\text{Im}(\varphi_{g,1}/\tau_g)$ by twice the width of the strip. Hence there is at most one stable shift of $\varphi_{g,1}$ for a given $n_\Omega$, and it exists only for certain values of the chemical potentials. On the second branch the conditions are $\text{Im}(1/\tau_g)<\text{Im}(\varphi_{g,1}/\tau_g)<0$. It is natural to conjecture that this result is trivially generalized to the four-charged solutions.

This is precisely analogous to the condition, $\Delta_{g,a}=[\Delta_{g,a}]_{\tau_g}$, $(a=1,2,3)$, for the $\mathcal{N}=4$ SYM index in \cite{Aharony:2021zkr}. Here $\Delta_{g,a}$ are the chemical potentials corresponding to our $\varphi_{g,a}$ and $[\Delta]_r$ is the periodic discontinuous function. This condition appeared to match the $\{1,N,r\}$ solution to the Bethe Ansatz equations. It is an open question to find the corresponding solution from the ABJM index which explains the condition, \eqref{multicond}.

\section{Accelerating and rotating black holes} \label{sec3}

\subsection{Black hole solutions}

We review the accelerating black holes by following \cite{Ferrero:2020twa}. The bosonic action of four-dimensional minimal gauged supergravity is
\begin{equation}
S\,=\,\frac{1}{16\pi{G}_{(4)}}\int{d}^4x\sqrt{-g}\left(R+6-F^2\right)\,,
\end{equation}
where $F=dA$. For the accelerating black hole solutions, the metric is given by
\begin{align} \label{minsol2}
ds^2\,=\,\frac{1}{H^2}\left\{-\frac{Q}{\Sigma}\left[dt-a\sin^2\theta\,d\phi\right]^2+\frac{\Sigma}{Q}dr^2+\frac{\Sigma}{P}d\theta^2\right. \,\, \quad & \notag \\
 \left.+\frac{P}{\Sigma}\sin^2\theta\left[adt-\left(r^2+a^2\right)d\phi\right]\right\}&\,,
\end{align}
where we have
\begin{align}
H\,=&\,1-\alpha\,r\cos\theta\,, \notag \\
\Sigma\,=&\,r^2+a^2\cos^2\theta\,, \notag \\
P\,=&\,1-2\alpha\,m\cos\theta+\big(\alpha^2\left(a^2+e^2+g^2\right)-a^2\big)\cos^2\theta\,, \notag \\
Q\,=&\,\big(r^2-2mr+a^2+e^2+g^2\big)\left(1-\alpha^2r^2\right)+\left(a^2+r^2\right)r^2\,.
\end{align}
The gauge field is 
\begin{equation}
A\,=\,-e\frac{r}{\Sigma}\left(dt-a\,\sin^2\theta\,d\phi\right)+g\frac{\cos\theta}{\Sigma}\left(adt-\left(r^2+a^2\right)d\phi\right)\,.
\end{equation}
The solution depends on five free parameters, $m$, $e$, $g$, $a$, and $\alpha$, which are loosely associated with mass, electric charge, magnetic charge, rotation and acceleration, respectively. We will consider the parameters to be
\begin{equation} \label{parasign}
m>0\,, \qquad \alpha,e,g,a\ge0\,.
\end{equation}
In the non-accelerating case, the solution reduces to the Kerr-Newman black hole solutions, \eqref{minsol}, up to some reparametrizations.

For the range of $0\le\theta\le\pi$, we choose, \cite{Ferrero:2020twa},
\begin{equation}
\theta_-\,=\,0\,, \quad \text{and} \quad \theta_+\,=\,\pi\,.
\end{equation}
Approaching $\theta=\theta_\pm$, the metric is
\begin{equation}
ds^2_{\theta,\phi}\,\approx\,\left[\frac{\Sigma}{PH^2}\right]_{\theta=\theta_\pm}\left[d\theta^2+P_\pm^2\left(\theta-\theta_\pm\right)^2d\phi^2\right]\,,
\end{equation}
where we define constants,
\begin{equation} \label{ppm}
P_\pm\,\equiv\,P(\theta_\pm)\,=\,1\pm2\alpha{m}+\alpha^2\left(a^2+e^2+g^2\right)-a^2\,.
\end{equation}
If there are conical deficits at both of the poles, they give rise to an orbifold known as a spindle, $\Sigma$, or equivalently, a weighted projective space. We set the ratio to be a rational number,
\begin{equation}
\frac{P_+}{P_-}\,=\,\frac{n_-}{n_+}\,,
\end{equation}
where $n_\pm\in\mathbb{N}$, and the period of $\phi$ is
\begin{equation}
\Delta\phi\,=\,\frac{2\pi}{n_+P_+}\,=\,\frac{2\pi}{n_-P_-}\,,
\end{equation}
and, from \eqref{ppm}, we find
\begin{equation}
\Delta\phi\,=\,\frac{\pi}{2m\alpha}\frac{n_--n_+}{n_-n_+}\,.
\end{equation}
The two-dimensional surface by $\theta$ and $\phi$ is the weighted projective space, $\mathbb{WCP}^1_{[n_-,n_+]}$.

The magnetic flux through the horizon is, \cite{Ferrero:2020twa},
\begin{equation} \label{mflux}
G_{(4)}Q_m\,\equiv\,\frac{1}{4\pi}\int_{r=r_+}F\,=\,g\frac{\Delta\phi}{2\pi}\,,
\end{equation}
and the electric flux is 
\begin{equation}
G_{(4)}Q_e\,\equiv\,\frac{1}{4\pi}\int_{r=r_+}*F\,=\,e\frac{\Delta\phi}{2\pi}\,.
\end{equation}
The angular momentum is defined by
\begin{equation}
G_{(4)}J\,=\,ma\left(\frac{\Delta\phi}{2\pi}\right)^2\,.
\end{equation}

\subsection{Supersymmetry}

We review the supersymmetry and thermodynamics of accelerating black holes, \cite{Ferrero:2020twa, Cassani:2021dwa}. 

From the integrability conditions for the Killing spinor equations, two constraints are found, \cite{Ferrero:2020twa},
\begin{align} \label{bpscond}
&0\,=\,2aeg\alpha+g^4\alpha^4+g^2\left(-1+\alpha^2+\left(a^2+2e^2\right)\alpha^4\right)+e^2\alpha^2\left(1+e^2\alpha^2+a^2\left(-1+\alpha^2\right)\right)\,, \notag \\
&0\,=\,2a^2g^2\alpha+2e^4\alpha^3+2g^4\alpha^3-2aeg\left(-1+a^2\alpha^2\right)-2e^2\alpha\left(a^2-2g^2\alpha^2\right) \notag \\
&-\alpha\left(m+a^2m\alpha^2\right)^2-\left(e^2+g^2\right)\alpha\left(1+a^2\alpha^2\right)\left(-1+e^2\alpha^2+g^2\alpha^2+a^2\left(-1+\alpha^2\right)\right)\,.
\end{align}
For the parameters in \eqref{parasign}, constraints are solved,
\begin{align}
m\,=&\,\frac{g}{\alpha}\,, \notag \\
a\,=&\,\frac{-eg\alpha+\left(e^2+g^2\right)\alpha^2\sqrt{1+\left(-1+e^2\right)\alpha^2-\left(e^2+g^2\right)\alpha^4}}{g^2\alpha^4+e^2\alpha^2\left(-1+\alpha^2\right)}\,.
\end{align}

If the radius of outer horizon, $r=r_+$, is a double root of $Q(r)$, the extremality condition with \eqref{bpscond} gives, \cite{Ferrero:2020twa},
\begin{equation}
a^3\alpha^2eg^3+a^2\alpha{g}^2\left(e-g\right)\left(e+g\right)-aeg^3+\alpha^3e^2\left(e^2+g^2\right)^2\,=\,0\,.
\end{equation}
The radius of the horizon for supersymmetric and extremal black holes is given by
\begin{equation}
r_+\,=\,\sqrt{\frac{a(m-ae)}{e+am\alpha^2}}\,,
\end{equation}

The magnetic flux, \eqref{mflux}, is given by
\begin{equation}
G_{(4)}Q_m\,=\,\frac{1}{4\pi}\int_\Sigma\,dA\,=\,\frac{n_--n_+}{4n_+n_-}\,.
\end{equation}
The Bekenstein-Hawking entropy of a black hole is given by
\begin{equation}
S_{BH}\,=\,\frac{1}{4G_{(4)}}\text{Area}\,=\,\frac{\sqrt{2}\sqrt{8n_-^2n_+^2\left(G_{(4)}Q_e\right)^2+n_-^2+n_+^2}-\left(n_-+n_+\right)}{n_-n_+}\frac{\pi}{4G_{(4)}}\,.
\end{equation}
The Euler number of the spindle horizon is given by
\begin{equation}
\chi(\Sigma)\,=\,\frac{1}{4\pi}\int_\Sigma\,R_2\text{vol}_2\,=\,\frac{n_-+n_+}{n_+n_-}\,,
\end{equation}
where $R_2$ is the Ricci scalar of the spindle.

Let us introduce a new parametrization of the parameters, \cite{Cassani:2021dwa},
\begin{equation}
e\,=\,\frac{bs}{\alpha^2c}\,, \quad g\,=\,\frac{s}{\alpha^2c}\,, \quad a\,=\,\frac{s}{\alpha}\,, \quad \Longleftrightarrow \quad b\,=\,\frac{e}{g}\,, \quad c\,=\,\frac{a}{g\alpha}\,, \quad s\,=\,a\alpha\,,
\end{equation}
and introduce $\rho$ from 
\begin{equation}
r_+\,\equiv\,\frac{s}{\alpha}\rho\,.
\end{equation}
We define the reduced chemical potentials
\begin{equation}
\omega\,\equiv\,\beta\left(\Omega-\Omega_*\right)\,, \qquad \varphi\,\equiv\,\beta\left(\Phi_e-\left(\Phi_e\right)_*\right)\,,
\end{equation}
where $\Omega$ is the angular velocity and $\Phi_e$ is the electrostatic potential and their extremal values are
\begin{equation}
\Omega_*\,=\,\frac{2}{\chi}\,, \qquad \left(\Phi_e\right)_*\,=\,1\,.
\end{equation}
We find
\begin{align}
\omega\,=&\,\frac{4\pi\left(\rho\mp{i}\right)s}{\chi\left(-1\mp{i}\rho\right)s+4G_{(4)}Q_m\left(\pm{i}+\rho{s}^2\right)}\,, \notag \\
\varphi\,=&\,\pm{i}\pi+\frac{\chi}{4}\omega\,.
\end{align}
The on-shell action can be expressed in terms of the complex chemical potentials,
\begin{equation} \label{oasusy2}
I\,=\,-S_{BH}-\omega{J}-\varphi{Q}_e\,=\,\pm\frac{1}{2iG_{(4)}}\left[\frac{\varphi^2}{\omega}+\left(G_{(4)}Q_m\right)^2\omega\right]\,.
\end{equation}
Note the change of the on-shell action, \eqref{oasusy2}, to the non-accelerating case, \eqref{oasusy}.

\subsection{Uplift to 11 dimensions}

We uplift the black hole solution of minimal gauged supergravity, \eqref{minsol2}, to a solution of eleven-dimensional supergravity. We employ the uplift formula for $U(1)^4$-gauged $\mathcal{N}=2$ supergravity, \cite{Cvetic:1999xp}, by taking all four $U(1)$ gauge fields to be equal.

The eleven-dimensional metric is given by
\begin{equation} \label{meleven2}
ds_{11}^2\,=\,ds_4^2+\frac{1}{g^2}\sum_{a=1}^4\left(d\mu_a^2+\mu_a^2\left(d\phi_a+gA\right)^2\right)\,,
\end{equation}
where $ds_4^2$ is the four-dimensional metric and $\mu_1^2+\mu_2^2+\mu_3^2+\mu_4^2=1$.

The four-form flux is given by
\begin{equation}
F_{(4)}\,=\,2g\epsilon_{(4)}-\frac{1}{2g^2}\sum_{a=1}^4d\left(\mu_a^2\right)\wedge{d}\phi_a\wedge*_4F\,,
\end{equation}
where $\epsilon_{(4)}$ is the volume form of the four-dimensional metric, $ds_4^2$, and $*_4$ is the Hodge dual with respect to that metric. From now on, we set $g=1/2$ to match the black hole solutions.

We calculate $F_{(7)}=*F_{(4)}$ and the six-form potential of $F_{(7)}=dC_{(6)}$ can be chosen to be
\begin{equation} \label{msixf2}
C_{(6)}\,=\,8d\left(\mu_2^2\right)\wedge{d}\left(\mu_3^2\right)\wedge{d}\phi_2\wedge{d}\phi_3\wedge{d}\phi_4\wedge{A}+\text{other\,\,terms}\,,
\end{equation}
where we only presented the term relevant to the discussion in the next section.

\subsection{The brane action}

Now we consider the M5-brane action at $r=r_+$, $\theta=\frac{\pi}{2}$, $\mu_1=0$, along the $\phi$ direction. The action of a M5-brane is given by, $e.g.$, \cite{Beccaria:2023cuo},
\begin{equation} \label{mdbi2}
S_\text{M5}\,=\,-\frac{1}{\left(2\pi\right)^5\ell_p^6}\int\left(d^6x\sqrt{-\text{det}\left(g_\text{M5}\right)}\mp{P}[C_{(6)}]\right)\,,
\end{equation}
where $g_\text{M5}$ is the induced metric on the M5-brane worldvolume, $P[C_{(6)}]$ is the pull-back of the six-form potential to the worldvolume, and the $\pm$ sign refers to a brane/anti-brane. Via the AdS/CFT correspondence, the Planck length, $\ell_p$, is related to the rank of the dual ABJM theory, $N$, by 
\begin{equation}
\frac{R_{S^7}^6}{\left(2\pi\right)^5\ell_p^6}\,=\,\frac{N}{\pi^3}\,.
\end{equation}
For the second term in the brane action, \eqref{mdbi2}, only the term of the six-form potential presented in \eqref{msixf2} contributes. It is a product of the integral of the integral of $8A$ on the $S^1$ in $AdS_4$ and the integral of $d\left(\mu_2^2\right)\wedge{d}\left(\mu_3^2\right)\wedge{d}\phi_2\wedge{d}\phi_3\wedge{d}\phi_4$ on the $S^5$ in $S^7$. The first integral is
\begin{equation}
8\int_{S^1}A\,=\,\frac{8ae}{r_+}\int_{S^1}d\phi\,=\,\frac{8ae}{r_+}\Delta\phi\,,
\end{equation}
and the second integral is $4\text{Vol}\left(S^5\right)=4\pi^3$.

Although the metric induced on the M5-brane from \eqref{meleven2} is not diagonal, the determinant is the product of the determinant of the $S^5$ metric and of $g_{\phi\phi}$ in \eqref{minsol2}. The integral of the determinant of $S^5$ with radius $2$ is $2^5\text{Vol}\left(S^5\right)=2^5\pi^3$. The component, $g_{\phi\phi}$, on the M5-brane is
\begin{equation}
g_{\phi\phi}\,=\,\left[\frac{r_+^2+a^2}{r_+}\right]^2\,,
\end{equation}
Choosing a sign for the square root, we obtain
\begin{equation} \label{metterm2}
\int\,d^6x\sqrt{-\text{det}\left(g_\text{M5}\right)}\,=\,-2^5\pi^3i\frac{r_+^2+a^2}{r_+}\Delta\phi\,,
\end{equation}
For the choice of the sign, the Lorentzian black hole is causally well-behaved and the contribution of \eqref{metterm2} to the path-integral measure, $\text{exp}(iS_\text{M5})$, is bounded in absolute value, \cite{Aharony:2021zkr}. The M5-brane action is
\begin{equation} 
S_\text{M5}\,=\,-\frac{N}{\pi^3}\left[-\frac{\pi^3}{2}i\frac{r_+^2+a^2}{r_+}\pm\frac{\pi^3}{2}\frac{ae}{r_+}\right]\Delta\phi\,,
\end{equation}
where the $\mp$ sign implies a brane/anti-brane.

Furthermore, the brane action can be expressed by
\begin{equation} \label{bactf2}
S_\text{M5}\,=\,\pm\pi{N}\left[\mp{i}\frac{1-\alpha^2r_+^2}{r_+}\frac{\varphi}{\omega}-\frac{1+a-\alpha^2r_+^2}{r_+}G_{(4)}Q_e\right]\,,
\end{equation}
which we checked numerically. We were not able to find an expression explicitly involving only the chemical potentials and charges, $e.g.$, without $r_+$, $a$, and $\alpha$. In appendix \ref{appC}, we show that the brane action reduces to the one for Kerr-Newman black holes in \eqref{bactf} for the non-accelerating case, $\alpha=0$. Also note the change of the on-shell action, \eqref{oasusy2}, to the non-accelerating case, \eqref{oasusy}.

\bigskip
\bigskip
\leftline{\bf Acknowledgements}
\noindent We would like to thank Davide Cassani and Dario Martelli for useful discussions. This work was supported by the National Research Foundation of Korea (NRF) grant
funded by the Korea government, Ministry of Science and ICT (RS-2026-25481139).

\bigskip
\leftline{\bf Use of artificial intelligence}
\noindent We used Claude Sonnet 5.5 (Anthropic) to assist with the derivations in section \ref{sgs},  equation \eqref{bactf2}, and appendix \ref{appC}. All results are checked by the author.

\appendix
\section{Rotating black holes with two electric charges} \label{appA}
\renewcommand{\theequation}{A.\arabic{equation}}
\setcounter{equation}{0} 

\subsection{Uplifting gauged STU supergravity to eleven dimensions}

Let us consider $U(1)^4$-gauged $\mathcal{N}=2$ supergravity in four dimensions, \cite{Duff:1999gh} and \cite{Cvetic:2000tb, Azizi:2016noi}. The field content is the metric, $ds_4^2$, three complex scalar fields from six real scalar fields, $\varphi_i$, $\chi_i$, $i=1,2,3$, and four $U(1)$ gauge fields, $A^\alpha$, $\alpha=1,2,3,4$. We introduce a parametrization for the scalar fields,
\begin{equation} \label{yparam}
Y_i^2\,=\,e^{\varphi_i}\,, \qquad \widetilde{Y}_i^2\,=\,e^{-\varphi_i}+\chi_i^2e^{\varphi_i}\,, \qquad b_i\,=\,\chi_ie^{\varphi_i}\,.
\end{equation}
For the details of the theory, we refer to section 2 of \cite{Azizi:2016noi}.

A solution of gauged STU supergravity can be uplifted to one of eleven-dimensional supergravity, \cite{Cremmer:1978km}. The uplift formula was constructed in \cite{Cvetic:2000tb, Azizi:2016noi}, and we review it by following section 5 of \cite{Azizi:2016noi}.

The eleven-dimensional metric is given by
\begin{align} \label{uplift11met}
ds_{11}^2\,=\,&\widetilde{\Xi}^\frac{1}{3}ds_4^2+\widetilde{\Xi}^\frac{1}{3}d\hat{s}_7^2 \notag \\
=\,&\widetilde{\Xi}^\frac{1}{3}ds_4^2+g^{-2}\widetilde{\Xi}^{-\frac{2}{3}}\Big[\sum_\alpha{Z}_\alpha(d\mu_\alpha^2+\mu_\alpha^2D\phi_\alpha^2)+2b_2b_3(\mu_1^2\mu_2^2D\phi_1D\phi_2-\mu_3^2\mu_4^2D\phi_3D\phi_4) \notag \\
&+2b_1b_3(\mu_1^2\mu_3^2D\phi_1D\phi_3-\mu_2^2\mu_4^2D\phi_2D\phi_4)+2b_1b_2(\mu_1^2\mu_4^2D\phi_1D\phi_4-\mu_2^2\mu_3^2D\phi_2D\phi_3) \notag \\
&+\frac{1}{2}b_1^2\Big((\mu_1d\mu_1+\mu_2d\mu_2)^2+(\mu_3d\mu_3+\mu_4d\mu_4)^2\Big) \notag \\
&+\frac{1}{2}b_2^2\Big((\mu_1d\mu_1+\mu_3d\mu_3)^2+(\mu_2d\mu_2+\mu_4d\mu_4)^2\Big) \notag \\
&+\frac{1}{2}b_3^2\Big((\mu_1d\mu_1+\mu_4d\mu_4)^2+(\mu_2d\mu_2+\mu_3d\mu_3)^2\Big)\Big]\,,
\end{align}
where $ds_4^2$ is the four-dimensional metric and $\mu_1^2+\mu_2^2+\mu_3^2+\mu_4^2=1$. We define $\widetilde{\Xi}$ by
\begin{align}
\widetilde{\Xi}\,=&\,Y_1^2Y_2^2Y_3^2\mu_1^4+Y_1\widetilde{Y}_2^2\widetilde{Y}_3^2\mu_2^4+\widetilde{Y}_1^2Y_2^2\widetilde{Y}_3^2\mu_3^4+\widetilde{Y}_1^2\widetilde{Y}_2^2Y_3^2\mu_4^4 \notag \\
&+(Y_2^2\widetilde{Y}_2^2+Y_3^2\widetilde{Y}_3^2)(Y_1^2\mu_1^2\mu_2^2+\widetilde{Y}_1^2\mu_3^2\mu_4^2) \notag \\
&+(Y_1^2\widetilde{Y}_1^2+Y_3^2\widetilde{Y}_3^2)(Y_2^2\mu_1^2\mu_3^2+\widetilde{Y}_2^2\mu_2^2\mu_4^2) \notag \\
&+(Y_1^2\widetilde{Y}_1^2+Y_2^2\widetilde{Y}_2^2)(Y_3^2\mu_1^2\mu_4^2+\widetilde{Y}_3^2\mu_2^2\mu_3^2)\,,
\end{align}
and $Z_\alpha$ by
\begin{equation}
W_\alpha\,=\,Z_\alpha-\mu_\alpha^2\,,
\end{equation}
where we have
\begin{align}
W_1\,=&\,\widetilde{Y}_2^2\widetilde{Y}_3^2\mu_2^2+\widetilde{Y}_1^2\widetilde{Y}_3^2\mu_3^2+\widetilde{Y}_1^2\widetilde{Y}_2^2\mu_4^2\,, \notag \\
W_2\,=&\,Y_2^2Y_3^2\mu_1^2+\widetilde{Y}_1^2Y_2^2\mu_3^2+\widetilde{Y}_1^2Y_3^2\mu_4^2\,, \notag \\
W_3\,=&\,Y_1^2Y_3^2\mu_1^2+Y_1^2\widetilde{Y}_2^2\mu_2^2+\widetilde{Y}_2^2Y_3^2\mu_4^2\,, \notag \\
W_4\,=&\,Y_1^2Y_2^2\mu_1^2+Y_1^2\widetilde{Y}_3^2\mu_2^2+Y_2^2\widetilde{Y}_3^2\mu_3^2\,.
\end{align}
The covariant derivative is given by
\begin{equation}
D\phi_\alpha\,\equiv\,d\phi_\alpha-gA_{(1)}^\alpha\,.
\end{equation}

The four-form flux is given by
\begin{equation} \label{uplift11flux}
\hat{F}_{(4)}\,=\,-2gU\epsilon_{(4)}+\hat{F}'_{(4)}+\hat{F}''_{(4)}+\hat{G}_{(4)}\,,
\end{equation}
where $\epsilon_{(4)}$ is the volume form of the four-dimensional metric, $ds_4^2$. In the first term, we define
\begin{equation}
U\,=\,Y_1^2\left(\mu_1^2+\mu_2^2\right)+\widetilde{Y}_1^2\left(\mu_3^2+\mu_4^2\right)+Y_2^2\left(\mu_1^2+\mu_3^2\right)+\widetilde{Y}_2^2\left(\mu_2^2+\mu_4^2\right)+Y_3^2\left(\mu_1^2+\mu_4^2\right)+\widetilde{Y}_3^2\left(\mu_2^2+\mu_3^2\right)\,.
\end{equation}
The second term is obtained from $\hat{F}'_{(4)}=d\hat{A}'_{(3)}$ where we have
\begin{equation}
\hat{A}'_{(3)}\,=\,\frac{1}{2}A_{\alpha\hat{\beta}\hat{\gamma}}d\mu_\alpha\wedge(d\phi_\beta-gA_{(1)}^\beta)\wedge(d\phi_\gamma-gA_{(1)}^\gamma)\,.
\end{equation}
We refer (4.19) of \cite{Azizi:2016noi} for the expressions of $A_{\alpha\hat{\beta}\hat{\gamma}}$. The third term is given by
\begin{equation}
\hat{F}''_{(4)}\,=\,-\frac{1}{2g^2}|\widetilde{W}|^{-2}\sum_\alpha{d}\mu_\alpha^2\wedge\left(d\phi_\alpha-gA_{(1)}^\alpha\right)\wedge{R}_\alpha\,,
\end{equation}
and we define
\begin{align}
R_1\,=\,&\widetilde{Y}_1^2\widetilde{Y}_2^2\widetilde{Y}_3^2[P_0*F_{(2)}^1+2b_1b_2b_3F_{(2)}^1]+\widetilde{Y}_1^2[P_1b_2b_3*F_{(2)}^2+b_1(P_0+2b_2^2b_3^2)F_{(2)}^2] \notag \\
&+\widetilde{Y}_2^2[P_2b_1b_3*F_{(2)}^3+b_2(P_0+2b_1^2b_3^2)F_{(2)}^3]+\widetilde{Y}_3^2[P_3b_1b_2*F_{(2)}^4+b_3(P_0+2b_1^2b_2^2)F_{(2)}^4]\,, \notag \\
R_2\,=\,&\widetilde{Y}_1^2Y_2^2Y_3^2[P_0*F_{(2)}^2+2b_1b_2b_3F_{(2)}^2]+\widetilde{Y}_1^2[P_1b_2b_3*F_{(2)}^1+b_1(P_0+2b_2^2b_3^2)F_{(2)}^1] \notag \\
&-Y_2^2[P_2b_1b_3*F_{(2)}^4+b_2(P_0+2b_1^2b_3^2)F_{(2)}^4]-Y_3^2[P_3b_1b_2*F_{(2)}^3+b_3(P_0+2b_1^2b_2^2)F_{(2)}^3]\,, \notag \\
R_3\,=\,&Y_1^2\widetilde{Y}_2^2Y_3^2[P_0*F_{(2)}^3+2b_1b_2b_3F_{(2)}^3]-Y_1^2[P_1b_2b_3*F_{(2)}^4+b_1(P_0+2b_2^2b_3^2)F_{(2)}^4] \notag \\
&+\widetilde{Y}_2^2[P_2b_1b_3*F_{(2)}^1+b_2(P_0+2b_1^2b_3^2)F_{(2)}^1]-Y_3^2[P_3b_1b_2*F_{(2)}^2+b_3(P_0+2b_1^2b_2^2)F_{(2)}^2]\,, \notag \\
R_4\,=\,&Y_1^2Y_2^2\widetilde{Y}_3^2[P_0*F_{(2)}^4+2b_1b_2b_3F_{(2)}^4]-Y_1^2[P_1b_2b_3*F_{(2)}^3+b_1(P_0+2b_2^2b_3^2)F_{(2)}^3] \notag \\
&-Y_2^2[P_2b_1b_3*F_{(2)}^2+b_2(P_0+2b_1^2b_3^2)F_{(2)}^2]+\widetilde{Y}_3^2[P_3b_1b_2*F_{(2)}^1+b_3(P_0+2b_1^2b_2^2)F_{(2)}^1]\,,
\end{align}
with
\begin{align}
&P_0\,\equiv\,1+b_1^2+b_2^2+b_3^2\,, \qquad \widetilde{W}\,\equiv\,P_0-2ib_1b_2b_3\,, \notag \\
&P_1\,\equiv\,1-b_1^2+b_2^2+b_3^2\,, \qquad P_2\,\equiv\,1+b_1^2-b_2^2+b_3^2\,, \qquad P_3\,\equiv\,1+b_1^2+b_2^2-b_3^2\,.
\end{align}
The fourth term is given by
\begin{align}
\hat{G}_{(4)}\,=&\,\frac{1}{2g}(2Y_1^{-1}*dY_1-\chi_1Y_1^4*d\chi_1)\wedge{d}(\mu_1^2+\mu_2^2) \notag \\
&+\frac{1}{2g}(2Y_2^{-1}*dY_2-\chi_2Y_2^4*d\chi_2)\wedge{d}(\mu_1^2+\mu_3^2) \notag \\
&+\frac{1}{2g}(2Y_3^{-1}*dY_3-\chi_3Y_3^4*d\chi_3)\wedge{d}(\mu_1^2+\mu_4^2)\,.
\end{align}

\subsection{The brane action}

The bosonic action of $U(1)^2$-gauged supergravity in four dimensions is, $e.g.$, \cite{BenettiGenolini:2023rkq},
\begin{align} \label{u2act}
S\,=\,\frac{1}{16\pi{G}_4}\int_{Y_4}\bigg[&\mathcal{R}\text{vol}_\mathcal{G}+\frac{1}{2}dX_1^2\wedge*dX_2^2-\frac{1}{2}\left(\varphi{X}_1^2\right)\wedge*d\left(\varphi{X}_1^2\right) \notag \\
&+\left(4+X_1^2+X_2^2\right)\text{vol}_\mathcal{G}-X_1^{-2}\left(\mathcal{F}_1\wedge*\mathcal{F}_1-\varphi{X}_1^2\mathcal{F}_1\wedge\mathcal{F}_1\right) \notag \\
&-X_2^{-2}\left(\mathcal{F}_2\wedge*\mathcal{F}_2+\varphi{X}_1^2\mathcal{F}_2\wedge\mathcal{F}_2\right)\bigg]\,,
\end{align}
where $\mathcal{G}$ is the metric, $\mathcal{F}_I=d\mathcal{A}_I$ with two $U(1)$ gauge fields, $\mathcal{A}_I$, $I=1,2$, $X_I$ are scalar fields, and $\varphi$ is a pseudo-scalar field. The scalar fields satisfy a relation,
\begin{equation}
X_2^2=X_1^{-2}+\varphi^2X_1^2\,.
\end{equation}
There is a $\mathbb{Z}_2$ automorphism from
\begin{equation}
\mathcal{F}_1\leftrightarrow\mathcal{F}_2\,, \qquad X_1\leftrightarrow{X}_2\,, \qquad \varphi{X}_1^2\leftrightarrow-\varphi{X}_1^2\,.
\end{equation}
If we have $X_1=X_2=1$ and $\mathcal{F}_1=\mathcal{F}_2=\mathcal{F}$, the action reduces to the one of minimal gauged supergravity, \eqref{minact}.

A black hole solution with a rotation and two electric charges, \cite{Chong:2004na, Cvetic:2005zi}, is 
\begin{align} \label{nminsol}
ds^2\,=&\,-\frac{\Delta_r\Delta_\theta}{B\Xi^2}dt^2+\sin^2\theta{B}\left(d\phi+a\Delta_\theta\frac{\Delta_r-\left(1+r_1r_2\right)\left(a^2+r_1r_2\right)}{BW\Xi^2}dt\right)^2 \notag \\
&+W\left(\frac{dr^2}{\Delta_r}+\frac{d\theta^2}{\Delta_\theta}\right)\,,
\end{align}
where we define
\begin{align}
r_i\,=&\,r+2ms_i^2\,, \notag \\
\Delta_r\,=&\,r^2+a^2-2mr+r_1r_2\left(r_1r_2+a^2\right)\,, \notag \\
\Delta_\theta\,=&\,1-a^2\cos^2\theta\,, \qquad W\,=\,r_1r_2+a^2\cos^2\theta\,, \qquad \Xi\,=\,1-a^2\,, \notag \\
B\,=&\,\frac{\Delta_\theta\left(r_1r_2+a^2\right)^2-a^2\sin^2\theta\Delta_r}{W\Xi^2}\,, 
\end{align}
and $s_i\,=\,\sinh\delta_i$, $c_i\,=\,\cosh\delta_i$, $i=1,2$. The scalar fields are
\begin{equation}
X_1^2\,=\,1+\frac{r_1\left(r_1-r_2\right)}{W}\,, \qquad \varphi\,=\,\frac{a\left(r_2-r_1\right)\cos\theta}{r_1^2+a^2\cos^2\theta}\,, \qquad X_2^2\,=\,1+\frac{r_2\left(r_2-r_1\right)}{W}\,,
\end{equation}
and the gauge fields are
\begin{align}
\mathcal{A}_1\,=\,\frac{2ms_2c_2r_1}{W\Xi}\left(\Delta_\theta\,dt-a\sin^2\theta\,d\phi\right)+\gamma_1dt\,, \notag \\
\mathcal{A}_2\,=\,\frac{2ms_1c_1r_2}{W\Xi}\left(\Delta_\theta\,dt-a\sin^2\theta\,d\phi\right)+\gamma_2dt\,,
\end{align}
where $\gamma_{1,2}$ are constant. The solution is in a frame that is non-rotating at infinity.

The largest root of $\Delta_r$, $r=r_+$, is where the outer horizon is. At constant $t$ and $r$ outside the horizon, the geometry is a sphere and closes off at $\theta=0,\pi$ with $\phi\sim\phi+2\pi$. We analytically continue to a Euclidean solution by a Wick rotation, $t=-it_\text{E}$. At the horizon, the geometry caps off smoothly, only with the identification, \cite{BenettiGenolini:2023rkq},
\begin{equation}
(t_\text{E},\phi)\,\sim\,(t_\text{E},\phi+2\pi)\,\sim\,(t_\text{E}+\beta,\phi-i\Omega\beta)\,,
\end{equation}
where the inverse temperature and angular velocity at the horizon are, respectively,
\begin{equation}
\beta\,=\,4\pi\frac{a^2+r_{1+}r_{2+}}{\Delta'_r(r_+)}\,, \qquad \Omega\,=\,a\frac{1+r_{1+}r_{2+}}{a^2+r_{1+}r_{2+}}\,,
\end{equation}
with $r_{i+}\equiv{r}_++2ms_i^2$.
The Bekenstein-Hawking entropy at the horizon is
\begin{equation}
S\,=\,\frac{\pi}{G_4}\frac{r_{1+}r_{2+}+a^2}{\Xi}\,.
\end{equation}
The electrostatic potentials are
\begin{equation}
\Phi_{e,1}\,=\,2ms_2c_2\frac{r_{1+}}{a^2+r_{1+}r_{2+}}\,, \qquad \Phi_{e,2}\,=\,2ms_1c_1\frac{r_{2+}}{a^2+r_{1+}r_{2+}}\,,
\end{equation}
where we fixed the gauge, $\gamma_1=-\Phi_{e,1}$ and $\gamma_2=-\Phi_{e,2}$. 

The supersymmetry condition for the black hole solutions, \eqref{nminsol}, is, \cite{Cvetic:2005zi, Chow:2013gba},
\begin{equation}
E\,=\,J+Q_{e,1}+Q_{e,2}\,, \qquad \Leftrightarrow \qquad a\,=\,\coth(\delta_1+\delta_2)-1\,.
\end{equation}

For the Euclidean solutions, the metric is generically complex, and the chemical potentials satisfy
\begin{equation}
\beta\left(1-\Phi_{e,1}-\Phi_{e,2}+\Omega\right)\,=\,\pm2\pi{i}\,,
\end{equation}
We define the ``reduced chemical potential",
\begin{equation}
\tau_g\,\equiv\,\beta\frac{\Omega-1}{2\pi{i}}\,, \qquad \varphi_{g1}\,\equiv\,\beta\frac{\Phi_{e,1}-1}{2\pi{i}}\,, \qquad \varphi_{g2}\,\equiv\,\beta\frac{\Phi_{e,2}-1}{2\pi{i}}\,,
\end{equation}
and they satisfy
\begin{equation} \label{condmult}
\tau_g-\varphi_{g1}-\varphi_{g2}\,=\,\pm1\,.
\end{equation}

For the BPS solutions, which are obtained by taking the extremality limit after imposing supersymmetry, the parameter, $m$, and the radius of the horizon are given by \cite{Cassani:2019mms},
\begin{equation}
m^2\,=\,\frac{\cosh^2\left(\delta_1+\delta_2\right)}{4e^{\delta_1+\delta_2}\sinh^3\left(\delta_1+\delta_2\right)c_1s_1c_2s_2}\,,
\end{equation}
\begin{equation}
r_*\,\equiv\,\frac{2ms_1s_2}{\cosh\left(\delta_1+\delta_2\right)}\,.
\end{equation}
The reduced chemical potentials are, \cite{Cassani:2019mms},
\begin{align} \label{cp1}
\tau_g\,=\,&-\frac{1}{2\pi{i}}\frac{16\pi}{\Theta}\left(e^{2\left(\delta_1+\delta_2\right)}-3\right)\Big[4\left(s_1c_1+s_2c_2\right)\sqrt{s_1s_2c_1c_2\left(s_1c_2+s_2c_1\right)e^{\delta_1+\delta_2}} \notag \\
&\mp4is_1s_2c_1c_2\left(c_1c_2+s_1s_2\right)e^{\delta_1+\delta_2}\Big]\,, \notag \\
\varphi_{g1}\,=&\,-\frac{1}{2\pi{i}}\frac{16\pi}{\Theta}\Big\{\sqrt{s_1s_2c_1c_2\left(s_1c_2+s_2c_1\right)e^{\delta_1+\delta_2}}[(e^{4\delta_2}-3)e^{2\delta_1}-4s_2c_2+2e^{-2\delta_1}] \notag \\
&\pm2is_2c_2[e^{2\left(\delta_1+\delta_2\right)}\left(c_1^2+s_1^2+2s_1c_1+2s_2c_2\right)-c_2^2-s_2^2-4s_2c_2]\Big\}\,,
\end{align}
where we define
\begin{align} \label{cp2}
\Theta\,=\,&e^{2\left(\delta_1+\delta_2\right)}\left(e^{4\delta_1}+e^{4\delta_2}-10\right)+6e^{4\left(\delta_1+\delta_2\right)}+e^{6\left(\delta_1+\delta_2\right)}-2\left(e^{-4\delta_1}+e^{-4\delta_2}\right) \notag \\
&-2[4\left(e^{4\delta_1}+e^{4\delta_2}\right)-7]+e^{-2\left(\delta_1+\delta_2\right)}[5\left(e^{4\delta_1}+e^{4\delta_2}\right)-3]\,,
\end{align}
and $\varphi_{g2}$ is obtained by switching $\delta_1$ and $\delta_2$ in $\varphi_{g1}$.

The on-shell action of the supersymmetric solutions, computed by holographic renormalization, is given by
\begin{equation}
I|_\text{SUSY}\,=\,\mp\frac{\pi}{G_4}\frac{\varphi_{g1}\varphi_{g2}}{\tau_g}\,.
\end{equation}

The action in \eqref{u2act}, is obtained by the $2+2$ truncation of $U(1)^4$-gauged $\mathcal{N}=2$ supergravity, \cite{Azizi:2016noi}, by identifying
\begin{align}
2+2: \qquad &\varphi_1\,=\,\varphi\,, \qquad \chi_1\,=\,\chi\,, \qquad \varphi_2\,=\,\varphi_3\,=\,\chi_2\,=\,\chi_3\,=\,0\,, \notag \\
&A_\mu^1\,=\,A_\mu^2\,=\,A_\mu\,, \qquad A_\mu^3\,=\,A_\mu^4\,=\,\widetilde{A}_\mu\,,
\end{align}
or, for the scalar fields in the parametrization of \eqref{yparam},
\begin{align}
Y_1\,&=\,Y\,, \qquad Y_2\,=\,Y_3\,=\,1\,, \notag \\
\widetilde{Y}_1\,&=\,\widetilde{Y}\,, \qquad \widetilde{Y}_2\,=\,\widetilde{Y}_3\,=\,1\,, \notag \\
b_1\,&=\,b\,, \qquad b_2\,=\,b_3\,=\,0\,.
\end{align}
Then, to employ the uplift formula of \cite{Azizi:2016noi}, which we have reviewed in appendix \ref{appA}, we identify our fields to theirs by
\begin{align}
Y\,=&\,X_1\,, \qquad \widetilde{Y}\,=\,X_2\,, \qquad b\,=\,\varphi{X}_1^2\,, \notag \\
A_\mu\,=&\,\mathcal{A}_{1\mu}\,, \qquad \widetilde{A}_\mu\,=\,\mathcal{A}_{2\mu}\,.
\end{align}
Accordingly, the uplifted eleven-dimensional metric and the four-form flux follow from \eqref{uplift11met} and \eqref{uplift11flux}, but we refrain from presenting the full expressions and will only present the necessary ones in the following. 

Now we study the M5-branes wrapping an $S^5$ inside the $S^7$ and an $S^1$ inside the $AdS_4$. They are at $r=r_+$, $\theta=\pi/2$, $\mu_2=0$, along the $\phi$ direction. For the metric given in \eqref{nminsol},
\begin{equation} \label{upmet}
ds_{11}\,=\,\widetilde{\Xi}^\frac{1}{3}ds_4^2+ds_{\widetilde{S}^7}^2\,,
\end{equation}
the relevant terms from the Hodge dual of the four-form flux are
\begin{align}
*\hat{F}''_{(4)}\,=&\,-\frac{1}{2g^2}|\widetilde{W}|^{-2}\left(\widetilde{Y}_1^2\widetilde{Y}_2^2\widetilde{Y}_3^2P_0\right)\tilde{*}_7\Big(d\left(\mu_\alpha^2\right)\wedge\left(d\phi_\alpha+A_{(1)}^\alpha\right)\Big)\wedge{F}_{(2)}^\alpha \notag \\
&+\text{other\,\,terms}\,,
\end{align}
where $\tilde{*}_7$ is the Hodge dual with respect to the metric on $\widetilde{S}^7$ in \eqref{upmet}. Thus we find
\begin{align}
*\hat{F}''_{(4)}\,=&\,-\frac{1}{2g^2}|\widetilde{W}|^{-2}\widetilde{Y}_1^2\widetilde{Y}_2^2\widetilde{Y}_3^2P_0\left(g^2\widetilde{\Xi}^{2/3}\right)^2\frac{1}{Z_1^2}\frac{1}{\mu_1^2}\sqrt{g_{\widetilde{S}^7}}\left(2\mu_1\right)d\mu_2\wedge{d}\mu_3\wedge{d}\phi_2\wedge{d}\phi_3\wedge{d}\phi_4\wedge{F}_{(2)}^1 \notag \\
&-\frac{1}{2g^2}|\widetilde{W}|^{-2}\widetilde{Y}_1^2Y_2^2Y_3^2P_0\left(g^2\widetilde{\Xi}^{2/3}\right)^2\frac{1}{Z_2^2}\frac{1}{\mu_2^2}\sqrt{g_{\widetilde{S}^7}}\left(2\mu_2\right)d\mu_1\wedge{d}\mu_3\wedge{d}\phi_1\wedge{d}\phi_3\wedge{d}\phi_4\wedge{F}_{(2)}^1 \notag \\
&+\text{other\,\,terms} \notag \\
=&\,+\frac{1}{4g^5}d\left(\mu_2^2\right)\wedge{d}\left(\mu_3^2\right)\wedge{d}\phi_2\wedge{d}\phi_3\wedge{d}\phi_4\wedge\mathcal{F}_1 \notag \\
&+\frac{1}{4g^5}d\left(\mu_1^2\right)\wedge{d}\left(\mu_3^2\right)\wedge{d}\phi_1\wedge{d}\phi_3\wedge{d}\phi_4\wedge\mathcal{F}_1 \notag \\
&+\text{other\,\,terms}\,,
\end{align}
where we employed
\begin{equation}
\sqrt{g_{\widetilde{S}^7}}\,=\,\frac{Z_1^{3/2}Z_3^{3/2}\mu_1\mu_2\mu_3\sqrt{Z_3\left(\mu_1^2+\mu_2^2\right)+\left(Z_1+b_1^2\left(\mu_1^2+\mu_2^2\right)\right)\left(\mu_3^2+\mu_4^2\right)}}{g^7\widetilde{\Xi}^{7/3}}\,.
\end{equation}
The corresponding terms of the six-form potential are given by
\begin{align}
C_{(6)}\,=&\,+\frac{1}{4g^5}d\left(\mu_2^2\right)\wedge{d}\left(\mu_3^2\right)\wedge{d}\phi_2\wedge{d}\phi_3\wedge{d}\phi_4\wedge\mathcal{A}_1 \notag \\
&+\frac{1}{4g^5}d\left(\mu_1^2\right)\wedge{d}\left(\mu_3^2\right)\wedge{d}\phi_1\wedge{d}\phi_3\wedge{d}\phi_4\wedge\mathcal{A}_1 \notag \\
&+\text{other\,\,terms}\,.
\end{align}

The terms of the six-form potential contribute to the M5-brane action by
\begin{equation}
\int_\text{M5}P[C_{(6)}]\,=\,\frac{1}{g^5}\int_{S^5}\mu_2d\mu_2\wedge\mu_3d\mu_3\wedge{d}\phi_2\wedge{d}\phi_3\wedge{d}\phi_4\wedge\int_{S^1}\mathcal{A}_{1\phi}d\phi\,=\,-\frac{4\pi^4}{g^5}\frac{mas_2c_2}{r_2\left(1-a^2\right)}\,.
\end{equation}
The contribution of the tension of the brane is from $\int_\text{M5}\sqrt{-\text{det}\left(g_\text{M5}\right)}$ and we find
\begin{align}
\sqrt{-\text{det}\left(g_\text{M5}\right)}\,=&\,\frac{1}{g^5}\sqrt{-g_{\phi\phi}^{(7)}}Y_1\mu_2\mu_3 \notag \\
=&\,\frac{1}{g^5}\sqrt{-\sin^2\theta{B}}X_1\mu_2\mu_3 \notag \\
=&\,-\frac{i}{g^5}\frac{r_1r_2+a^2}{\sqrt{r_1r_2}\left(1-a^2\right)}\sqrt{1+\frac{r_1\left(r_1-r_2\right)}{r_1r_2}}\mu_2\mu_3\,.
\end{align}

The action of the M5-brane along $\mu_2=0$ is given by
\begin{align}
S_\text{M5}\,=&\,-\frac{Ng^6}{\pi^3}\int_\text{M5}\left(d^6x\sqrt{-\text{det}\left(g_\text{M5}\right)}\mp{P}[C_{(6)}]\right) \notag \\
=&\,-\frac{Ng^6}{\pi^3}\frac{2\pi^4}{g^5}\left[-i\frac{r_1r_2+a^2}{\sqrt{r_1r_2}\left(1-a^2\right)}\sqrt{1+\frac{r_1\left(r_1-r_2\right)}{r_1r_2}}\pm\frac{2mas_2c_2}{r_2\left(1-a^2\right)}\right] \notag \\
=&\,\pm2\pi{N}g\frac{2}{e^{2\left(\delta_1+\delta_2\right)}-3}\left[-1\pm{i}\sqrt{e^{\delta_1+\delta_2}\sinh\left(\delta_1+\delta_2\right)\frac{\sinh\left(2\delta_1\right)}{\sinh\left(2\delta_2\right)}}\right] \notag \\
=&\,\pm2\pi{N}g\frac{\Phi_{e,1}-1}{\Omega-1}\,=\,\pm2\pi{N}g\frac{\varphi_{g1}}{\tau_g}\,,
\end{align}
where, in the last equality, we employed \eqref{cp1} and \eqref{cp2}. As for the black hole solutions in minimal gauged supergravity, the brane action is expressed in terms of the reduced chemical potentials.

Similarly, the action of M5-branes along $\mu_1=0$ is given by
\begin{align}
S_\text{M5}\,=&\,-\frac{Ng^6}{\pi^3}\int_\text{M5}\left(d^6x\sqrt{-\text{det}\left(g_\text{M5}\right)}\mp{P}[C_{(6)}]\right) \notag \\
=&\,-\frac{Ng^6}{\pi^3}\frac{2\pi^4}{g^5}\left[-i\frac{r_1r_2+a^2}{\sqrt{r_1r_2}\left(1-a^2\right)}\sqrt{1+\frac{r_2\left(r_2-r_1\right)}{r_1r_2}}\pm\frac{2mas_1c_1}{r_1\left(1-a^2\right)}\right] \notag \\
=&\,\pm2\pi{N}g\frac{2}{e^{2\left(\delta_1+\delta_2\right)}-3}\left[-1\pm{i}\sqrt{e^{\delta_1+\delta_2}\sinh\left(\delta_1+\delta_2\right)\frac{\sinh\left(2\delta_2\right)}{\sinh\left(2\delta_1\right)}}\right] \notag \\
=&\,\pm2\pi{N}g\frac{\Phi_{e,2}-1}{\Omega-1}\,=\,\pm2\pi{N}g\frac{\varphi_{g2}}{\tau_g}\,.
\end{align}
This agrees with the previous result of the M5-brane action along $\mu_2=0$.

In appendix D of \cite{Aharony:2021zkr}, another kind of D3-brane wrapping an $S^3$ in the $AdS_5$ and $S^1$ in the $S^5$ was also studied. However, this kind of brane appears to exist only in the $AdS_5\times{S}^5$ background because the dimensionality of $AdS_5$ and $S^5$ are equal. In particular, the flux supporting such a brane has a part from $F_{(5)}$ and also from $*F_{(5)}$. On the other hand, for the $AdS_4\times{S}^7$ background we are studying here, for an analogous brane, a part of the flux is from $F_{(4)}$ and also from $*F_{(4)}$, and the dimensionality does not match to combine them.

\section{Seven-dimensional black holes} \label{appB}
\renewcommand{\theequation}{B.\arabic{equation}}
\setcounter{equation}{0} 

The field content of $U(1)^2$-gauged supergravity in seven dimensions is the metric, a three-form potential, $A_{(3)}$, two Abelian gauge fields, $A_{(I)}$, $I=1,2$, in the Cartan of $SO(5)$, and two real scalar fields, $\varphi_i$, $i=1,2$. The bosonic Lagrangian is, \cite{Pernici:1984xx, Liu:1999ai},
\begin{align}
\mathcal{L}_7\,&=\,R\star1-\frac{1}{2}\sum_{i=1}^2\star{d}\varphi_i\wedge{d}\varphi_i-\frac{1}{2}\sum_{I=1}^2L_I^{-2}\star{F}_{(2)}^I\wedge{F}_{(2)}^I-\frac{1}{2}\left(L_1L_2\right)^2\star{F}_{(4)}\wedge{F}_{(4)} \notag \\
&-2g^2[\left(L_1L_2\right)^{-4}-8L_1L_2-4L_1^{-1}L_2^{-2}-4L_1^{-2}L_2^{-1}]\star1 \notag \\
&-gF_{(4)}\wedge{A}_{(3)}+F_{(2)}^1\wedge{F}_{(2)}^2\wedge{A}_{(3)}\,,
\end{align}
where we have
\begin{align}
F_{(2)}^I\,&=\,dA_{(1)}^I\,, \qquad \qquad \quad F_{(4)}\,=\,dA_{(3)}\,, \notag \\
L_1\,&=\,e^{-\frac{1}{\sqrt{2}}\varphi_1-\frac{1}{\sqrt{10}}\varphi_2}\,, \qquad L_2\,=\,e^{\frac{1}{\sqrt{2}}\varphi_1-\frac{1}{\sqrt{10}}\varphi_2}\,,
\end{align}
and $g$ is the gauge coupling constant. The first-order self-duality condition for the four-form flux is given by
\begin{equation}
F_{(3)}\,=\,dA_{(2)}-\frac{1}{2}A_{(1)}^1\wedge{d}A_{(1)}^2-\frac{1}{2}A_{(1)}^2\wedge{A}_{(1)}^1\,,
\end{equation}
and the self-duality equation is
\begin{equation}
\left(L_1L_2\right)^2\star{F}_{(4)}\,=\,-2gA_{(3)}-F_{(3)}\,.
\end{equation}

The single-rotation two-charge black hole solution is given by, \cite{Chong:2004dy, Cvetic:2005zi, Hosseini:2018dob},
\begin{align}
ds^2\,=&\,\left(H_1H_2\right)^{1/5}\left(-\frac{V}{H_1H_2B}r^2dt^2+B\left(\sigma+fdt\right)^2+\frac{dr^2}{V}+r^2ds_{\mathbb{CP}^2}^2\right)\,, \notag \\
A_{(1)}^I\,=&\,\frac{2ms_I}{\rho^4\Xi{H}_I}\left(\alpha_I\Xi_-dt+\beta_I\sigma\right)\,, \notag \\
A_{(2)}\,=&\,\frac{mas_1s_2}{\rho^4\Xi_-}\left(\frac{1}{H_1}+\frac{1}{H_2}\right)dt\wedge\sigma\,, \quad A_{(3)}\,=\,\frac{2mas_1s_2}{\rho^2\Xi\Xi_-}\sigma\wedge{J}\,, \notag \\
L_I\,=&\,\left(H_1H_2\right)^{2/5}H_I^{-1}\,, \quad H_I\,=\,1+\frac{2ms_I^2}{\rho^4}\,, \quad \rho\,=\,\sqrt{\Xi}r\,, \notag \\
\alpha_1\,=&\,c_1-\frac{1}{2}\left(1-\Xi_+^2\right)\left(c_1-c_2\right)\,, \quad \alpha_2\,=\,c_2+\frac{1}{2}\left(1-\Xi_+^2\right)\left(c_1-c_2\right)\,, \notag \\
\beta_1\,=&\,-a\alpha_2\,, \quad \beta_2\,=\,-a\alpha_1\,, \quad \Xi_\pm\,=\,1\pm{ag}\,, \quad \Xi\,=\,1-a^2g^2\,, \notag \\
s_I\,\equiv&\,\sinh\delta_I\,, \quad c_I\,\equiv\,\cosh\delta_I\,,
\end{align}
where $\sigma=d\phi+\mathcal{B}$ and $\frac{1}{2}d\mathcal{B}$ is the K\"ahler form on $\mathbb{CP}^2$. The metric functions are given by
\begin{equation} \label{sevenf}
V\,=\,\frac{Y}{\Xi\rho^6}\,, \quad B\,=\,\frac{f_1}{H_1H_2\Xi^2\rho^4}\,, \quad f\,=\,-\frac{2f_2\Xi_-}{f_1}\,,
\end{equation}
and they are functions of the radial coordinate, $r$. We also define
\begin{align}
f_1\,=&\,\Xi\rho^6H_1H_2-\frac{4\Xi_+^2m^2a^2s_1^2s_2^2}{\rho^4}+\frac{1}{2}ma^2[4\Xi_+^2-2c_1c_2\left(\Xi_+^4-1\right)+\left(c_1^2+c_2^2\right)\left(\Xi_+^2-1\right)^2]\,, \notag \\
f_2\,=&\,-\frac{1}{2}g\Xi_+\rho^6H_1H_2+\frac{1}{4}ma[2c_1c_2\left(\Xi^4+1\right)-\left(c_1^2+c_2^2\right)\left(\Xi_+^2-1\right)]\,, \notag \\
Y\,=&\,g^2\rho^8H_1H_2+\Xi\rho^6+\frac{1}{2}ma^2[4\Xi_+^2-2c_1c_2\left(\Xi_+^4-1\right)+\left(c_1^2+c_2^2\right)\left(\Xi_+^2-1\right)^2] \notag \\
&-\frac{1}{2}m\rho^2[4\Xi+2c_1c_2a^2g^2\left(3a^2g^2+8ag+6\right)-\left(c_1^2+c_2^2\right)a^2g^2\left(ag+2\right)\left(3ag+2\right)]\,.
\end{align}
There are two free parameters due to the constrains,
\begin{align}
&e^{\delta_1+\delta_2}\,=\,1-\frac{2}{3ag}\,, \notag \\
&m\,=\,\frac{128e^{\delta_1+\delta_2}\left(3e^{\delta_1+\delta_2}-1\right)^3}{729g^4\left(e^{2\delta_1}-1\right)\left(e^{2\delta_2}-1\right)\left(e^{\delta_1+\delta_2}+1\right)^2\left(e^{\delta_1+\delta_2}-1\right)^4}\,, 
\end{align}
where the first one is from the BPS conditions and the second one is to avoid naked closed timelike curves (CTCs). The radius of horizon from $V(r_0)=0$ is 
\begin{equation}
r_0^2\,=\,\frac{16}{3g^2\left(e^{\delta_1+\delta_2}+1\right)\left(3e^{\delta_1+\delta_2}-5\right)}\,.
\end{equation}
The metric on $\mathbb{CP}^2$ is the standard Fubini-Study metric,
\begin{equation}
ds_{\mathbb{CP}^2}^2\,=\,d\xi^2+\frac{1}{4}\sin^2\xi\left(\sigma_1^2+\sigma_2^2+\cos^2\xi\sigma_3^2\right)\,,
\end{equation}
where $\sigma_i$, $i=1,2,3$, are left-invariant one-forms on $SU(2)$ satisfying $d\sigma_i\,=\,-\frac{1}{2}\epsilon_{ijk}\sigma_j\wedge\sigma_k$.

We consider the equal charge case, $\delta\equiv\delta_1=\delta_2$, with $s\equiv{s}_1=s_2$ and $c\equiv{c}_1=c_2$. We also set $g=1$. For the BPS solutions, the reduced chemical potentials are defined to be, \cite{Cassani:2019mms},
\begin{equation}
\omega\,=\,\beta\left(\Omega-\Omega^*\right)\,, \qquad \varphi\,=\,\beta\left(\Phi-\Phi^*\right)\,,
\end{equation}
where $\Omega$ and $\Phi$ are the angular momentum and electrostatic potential, respectively. The BPS values of the chemical potentials are
\begin{equation}
\Omega^*\,=\,1\,, \quad \Phi^*\,=\,1\,, \quad \beta\,\rightarrow\,\infty\,.
\end{equation}
The on-shell action is given by
\begin{equation}
I\,=\,-\frac{\pi^3}{128}\frac{\varphi^4}{\omega^3}\,.
\end{equation}

If the extremal limit is taken, the reduced chemical potentials are, \cite{Cassani:2019mms},
\begin{align} \label{cp11}
\omega^*\,=\,-&\frac{6\pi}{\left(c+8s\right)\Theta}\sqrt{2\left(16cs+c^2+s^2+1\right)}\left(c-4s\right)\left[c\Big(\sqrt{3}\mp\sqrt{-8\tanh\delta-1}\Big)\right. \notag \\
&+4s\left(\pm\sqrt{-8\tanh\delta-1}+2\sqrt{3}\right)\Big]\,, \notag \\
\varphi^*\,=\,-&\frac{64\pi}{\Theta{e}^\delta\sqrt{2\left(c^2+16cs+s^2+1\right)}}\Big(\sqrt{3}e^{-\delta}\left(c+8s\right)\pm9cs\sqrt{-8\tanh\delta-1}\Big)\,,
\end{align}
where we define
\begin{align} \label{cp22}
\Theta\,=&\,8cs\left(\pm\sqrt{-3\left(8\tanh\delta+1\right)}-18\right)+\left(\pm23\sqrt{-3\left(8\tanh\delta+1\right)}-9\right)\left(c^2+s^2\right) \notag \\
&-9\left(\pm\sqrt{-3\left(8\tanh\delta+1\right)}+1\right)\,.
\end{align}
The BPS chemical potentials satisfy
\begin{equation}
3\omega^*-2\varphi^*\,=\,\mp2\pi{i}\,.
\end{equation}
and the on-shell action is
\begin{equation}
I^*\,=\,-\frac{\pi^3}{128}\frac{\left(\varphi^*\right)^4}{\left(\omega^*\right)^3}\,.
\end{equation}

We employ the uplift formula for $U(1)^2$-gauged supergravity in seven dimensions to eleven-dimensional supergravity, \cite{Cvetic:1999xp}. The eleven-dimensional metric is given by
\begin{equation} \label{stoe}
ds_{11}^2\,=\,\tilde{\Delta}^{1/3}ds_7^2+g^{-2}\tilde{\Delta}^{-2/3}\Big(X_0^{-1}d\mu_0^2+\sum_{i=1}^2X_i^{-1}\left(d\mu_i^2+\mu_i^2\left(d\phi_i+gA_{(1)}^i\right)^2\right)\Big)\,, 
\end{equation}
where $ds_7^2$ is the seven-dimensional metric and $\mu_0^2+\mu_1^2+\mu_2^2=1$. We define $X_0\equiv\left(X_1X_2\right)^{-2}$ and
\begin{equation}
\tilde{\Delta}\,=\,\sum_{\alpha=0}^2X_\alpha\mu_\alpha^2\,.
\end{equation}
The Hodge dual of four-form flux is given by
\begin{align}
*F_{(4)}\,=\,&2g\sum_{\alpha=0}^2\left(X_\alpha^2\mu_\alpha^2-\tilde{\Delta}X_\alpha\right)\epsilon_{(7)}+g\tilde{\Delta}X_0\epsilon_{(7)}+\frac{1}{2g}\sum_{\alpha=0}^2X_\alpha^{-1}*_7dX_\alpha\wedge{d}\left(\mu_\alpha^2\right) \notag \\
&+\frac{1}{2g^2}\sum_{i=1}^2X_i^{-2}d\left(\mu_i^2\right)\wedge\left(d\phi_i+gA_{(1)}^i\right)\wedge*_7F_{(2)}^i\,,
\end{align}
where $\epsilon_{(7)}$ is the volume form of the seven-dimensional metric, $ds_7^2$, and $*_7$ is the Hodge dual with respect to that metric.

Now we study the M2-branes wrapping an $S^2$ inside the $S^4$ and an $S^1$ inside the $AdS_7$. They are at $r=r_+$ and $\mu_1=0$ along the $\phi$ direction. For the metric given in \eqref{stoe},
\begin{equation}
ds_{11}^2\,=\,\tilde{\Delta}^{1/3}ds_7^2+ds_{\tilde{S}^4}^2\,, 
\end{equation}
the three-form potential $C_{(3)}$ can be chosen to be
\begin{equation}
C_{(3)}\,=\,\frac{1}{2g^2}\frac{1}{\mu_0}d\left(\mu_2^2\right)\wedge{d}\phi_2\wedge{A}_{(1)}^1+\text{other terms}\,,
\end{equation}
where we only presented the term relevant to the discussion in the following. We employed
\begin{align}
\sqrt{g_{\tilde{S}^4}}\,=&\,\left(\frac{1}{g^2\tilde{\Delta}^{2/3}}\right)^2\frac{\mu_1\mu_2\left(X_0\mu_0^2+X_1\mu_1^2+X_2\mu_2^2\right)^{1/2}}{\mu_0\left(X_0X_1^2X_2^2\right)^{1/2}} \notag \\
=&\,\left(\frac{1}{g^2\tilde{\Delta}^{2/3}}\right)^2\frac{\mu_1\mu_2}{\mu_0}\tilde{\Delta}^{1/2}\,.
\end{align}

The terms of the three-form potential contribute to the M2-brane action by
\begin{equation}
\int_\text{M2}P[C_{(3)}]\,=\,\frac{1}{g^2}\int_{S^2}\frac{1}{\mu_0}\mu_2d\mu_2\wedge{d}\phi_2\wedge\int_{S^1}A_{(1)\phi}^1d\phi\,.
\end{equation}
The contribution of the tension of the brane is from $\int_\text{M2}\sqrt{-\text{det}\left(g_\text{M2}\right)}$ and we find
\begin{equation}
\sqrt{-\text{det}\left(g_\text{M2}\right)}\,=\,i\sqrt{B}\,,
\end{equation}
where $B$ is defined in \eqref{sevenf}.

The action of the M2-brane is given by
\begin{align}
S_\text{M2}\,=&\,\frac{1}{\left(2\pi\right)^3\ell_p^3}\int\left(d^3x\sqrt{-\text{det}\left(g_\text{M2}\right)}\pm{P}[C_{(3)}]\right) \notag \\
=&\,\frac{N}{8\pi^2}\left[-4\pi^3i\sqrt{B}\pm4\pi^3A_{(1)\phi}^1\right] \notag \\
=&\,\frac{\pi}{2}N\left(4\frac{3+i\sqrt{3\left(1+8\tanh\delta\right)}}{15-9e^{2\delta}}\right) \notag \\
=&\,\frac{\pi}{2}N\left(\frac{1}{2}\frac{\varphi^*}{\omega^*}\right) \notag \\
=&\,\frac{\pi}{4}N\frac{\varphi^*}{\omega^*}\,,
\end{align}
where, in the last equality, we employed \eqref{cp11} and \eqref{cp22}. We also have
\begin{equation}
\left(\frac{R_{S^4}}{\ell_p}\right)^3\,=\,\pi{N}\,,
\end{equation}
for $AdS_7\times{S}^4$ with $R_{S^4}=1$. The brane action is expressed in terms of the reduced chemical potentials.

\section{Reducing brane action to non-accelerating case} \label{appC}
\renewcommand{\theequation}{C.\arabic{equation}}
\setcounter{equation}{0} 

In this appendix, for the non-accelerating case, $\alpha=0$, we show that the brane action for accelerating black holes, \eqref{bactf2}, reduces to the one for Kerr-Newman black holes in \eqref{bactf}.

For the non-accelerating case, $\alpha=0$, we consider the spherical horizon with $\chi=2$ and $n_\pm=1$. From appendix B of \cite{Ferrero:2020twa}, we have
\begin{equation}
g\,=\,0\,, \qquad m\,=\,(1+a)e\,,
\end{equation}
and
\begin{equation}
e\,=\,\sqrt{a}(1+a)\,.
\end{equation}
We also find
\begin{equation}
G_{(4)}Q_e\,=\,\frac{e}{1-a^2}\,,
\end{equation}
and
\begin{equation}
\Delta\phi\,=\,\frac{2\pi}{1-a^2}\,.
\end{equation}

The $Q$ function is expressed by a sum of squares,
\begin{equation}
Q\,=\,\left(r^2-a\right)^2+\Big((1+a)r-e\Big)^2\,.
\end{equation}
At the horizon, $Q\left(r_+\right)=0$, we find
\begin{equation} \label{erp}
e\,=\,(1+a)r_+\mp{i}\left(r_+^2-a\right)\,.
\end{equation}
Let us express the Q function by
\begin{equation}
Q\,=\,X^2+Y^2\,=\,\left(X+iY\right)\left(X-iY\right)\,,
\end{equation}
where we have
\begin{equation}
X\,=\,r^2-a\,, \qquad Y\,=\,(1+a)r-e\,.
\end{equation}
Then, at the horizon, $Q(r_+)=0$, we have
\begin{equation}
Y\,=\,\pm{i}X\,.
\end{equation}
Using
\begin{equation}
X'\,=\,2r\,, \qquad Y'\,=\,1+a\,, 
\end{equation}
the derivative of Q is given by
\begin{align} \label{dQ}
Q'(r_+)\,=&\,2XX'+2YY'\,=\,4r_+X+2(1+a)Y \notag \\
=&\,4r_+X\pm2i(1+a)X\,=\,X\Big(4r_+\pm2i(1+a)\Big) \notag \\
=&\,\left(r_+^2-a\right)\Big(4r_+\pm2i(1+a)\Big)\,.
\end{align}

For the non-accelerating case, $\alpha=0$, we have the electrostatic potential, the angular momentum at the horizon and at infinity and the inverse temperature given by, respectively, \cite{Cassani:2021dwa},
\begin{align}
\Phi_e\,=&\,\frac{er_+}{r_+^2+a^2}\,, \qquad \qquad \Omega_H\,=\,\frac{2\pi}{\Delta\phi}\frac{a}{r_+^2+a^2}\,, \notag \\
\Omega_\infty\,=&\,-\frac{2\pi}{\Delta\phi}\frac{a}{1-a^2}\,, \quad \quad \,\,\, \beta\,=\,\frac{4\pi\left(r_+^2+a^2\right)}{Q'(r_+)}\,,
\end{align}
and find
\begin{equation}
\Omega\,=\,\Omega_H-\Omega_\infty\,=\,\frac{a\left(r_+^2+1\right)}{r_+^2+a^2}\,.
\end{equation}
The chemical potentials for the supersymmetric and extremal case are
\begin{equation}
\Omega_*\,=\,1\,, \qquad \left(\Phi_e\right)_*\,=\,1\,.
\end{equation}
Using \eqref{erp} and \eqref{dQ}, the reduced chemical potentials are given by
\begin{align}
\omega\,\equiv&\,\beta\left(\Omega-1\right)\,=\,-\frac{2\pi(1-a)}{2r_+\pm{i}(1+a)}\,, \notag \\
\varphi\,\equiv&\,\beta\left(\Phi_e-1\right)\,=\,\frac{2\pi\left(a\mp{i}r_+\right)}{2r_+\pm{i}(1+a)}\,,
\end{align}
and find
\begin{equation}
\frac{\varphi}{\omega}\,=\,\frac{a\mp{i}r_+}{a-1}\,, \qquad \varphi-\frac{\omega}{2}\,=\,\mp{i}\pi\,.
\end{equation}
Finally, we find that the brane action for an accelerating black hole, \eqref{bactf2}, reduces to the one for a Kerr-Newman black hole, \eqref{bactf}, 
\begin{align}
S_\text{M5}\,=&\,\pm\pi{N}\left[\mp{i}\frac{1}{r_+}\frac{\varphi}{\omega}-\frac{1+a}{r_+}G_{(4)}Q_e\right] \notag \\
=&\,\pm\pi{N}\frac{\varphi}{\omega}\,.
\end{align}

\bibliographystyle{JHEP}
\bibliography{20260321_ref}

\end{document}